\documentclass[twocolumn,aps,prl,amsmath,amssymb,superscriptaddress]{revtex4-2}

\usepackage{newtxtext}
\usepackage{newtxmath}
\usepackage{microtype}
\usepackage{wasysym}
\usepackage{graphicx}
\usepackage{dcolumn}
\usepackage{bm}
\usepackage{multirow}
\usepackage{amssymb}
\usepackage{graphicx}
\usepackage{xcolor}
\definecolor{TighnariBrown}{RGB}{141,87,41}
\definecolor{TighnariGreen}{RGB}{40,114,70}
\definecolor{TighnariYellow}{RGB}{247,191,99}
\definecolor{PRLBlue}{RGB}{46,48,146}
\usepackage[
    colorlinks,
    citecolor=PRLBlue,
    linkcolor=PRLBlue,
    urlcolor=PRLBlue,
]{hyperref}
\usepackage{indentfirst}
\usepackage{cases}

\begin{document}
\title{Finite-time Scaling with Arbitrary Driving Rates: Bridging the Kibble-Zurek and De~Grandi-Gritsev-Polkovnikov Limits}
\author{Shuai Yin}
\email{yinsh6@mail.sysu.edu.cn}
\affiliation{School of Physics, Sun Yat-sen University, Guangzhou 510275, China}
\affiliation{Guangdong Provincial Key Laboratory of Magnetoelectric Physics and Devices, Sun Yat-sen University, Guangzhou 510275, China}

\date{\today}

\begin{abstract}
The pursuit of a universal description for nonequilibrium critical dynamics in quantum many-body systems stands as a central frontier in modern statistical physics. For driven critical dynamics starting far from the critical point, the well-known Kibble-Zurek (KZ) scaling holds only when the driving rate lies below an upper bound. Here we study driven dynamics restricted to the critical region, and show that robust dynamic scaling behavior exists for arbitrary driving rates. We develop a generalized finite-time scaling (FTS) framework, which provides a unified understanding on the driven dynamics for the full range of quench rates, bridging the KZ scaling in the slow-driving regime and the De~Grandi-Gritsev-Polkovnikov (DGP) scaling in the sudden-quench limit. We verify this unified FTS form through numerical simulations in both quantum critical and tricritical points. The good agreement between theoretical predictions and numerical results confirms the generality of our theory. Our work establishes a universal theory for nonequilibrium critical dynamics spanning the full range of driving rates, with broad implications for quantum quench experiments and out-of-equilibrium statistical mechanics.
\end{abstract}

\maketitle

{\bf Introduction}---Unlike equilibrium states, which are uniquely determined, nonequilibrium states can manifest in remarkably diverse configurations~\cite{Rigol2016review,Mitra2018arcmp,Dziarmaga2010review,Polkovnikov2011rmp,delcamporev}. Establishing a unified theoretical framework to describe nonequilibrium dynamics has long remained a fundamental open problem in statistical physics. Among various topics in nonequilibrium physics, the dynamics accompanying phase transitions has attracted intensive attention. The interplay between universal critical properties and time-dependent dynamics yields a wealth of universal dynamic behaviors. Two paradigmatic schemes have been investigated: near-adiabatic ramps across the critical point ~\cite{Kibble1976,Zurek1985,Zoller2005prl,Dziarmaga2005prl,Fischer2006prl,Sen2008prl,Du2023,Navon2015science,Lamporesi2013,Lin2014natphy,delCampo2020prl,delcampo2023prl,Navon2015science,Quan2020prl,Lifux2025prl,doi:10.1126/science.abq6753} and sudden instantaneous quenches starting from the critical point~\cite{Polkovnikov2010prb,Dengshusa2011prb}. These correspond respectively to the limits of slow and infinitely fast driving, each governed by a well-established theory.

For near-adiabatically driven critical dynamics, the renowned Kibble–Zurek (KZ) mechanism~\cite{Dziarmaga2010review,Polkovnikov2011rmp,delcamporev,Kibble1976,Zurek1985}, first proposed by Kibble in cosmology~\cite{Kibble1976} and later extended to condensed-matter physics by Zurek~\cite{Zurek1985}, provides a framework for describing topological defect formation, as well as the resultant scaling behavior of defect density with external driving rates. When the relevant parameter $\lambda$ is linearly tuned beginning with a value $\lambda_i$, to drive the system across its critical point $\lambda_c$ at a rate $R$, following the relation $\lambda(t)-\lambda_c =Rt$, the KZ mechanism shows that the density of topological defects $n$ satisfies 
\begin{equation}
n\propto R^{d/r},
\label{kz}
\end{equation}
in which $d$ is the dimension of the system, $r=z+1/\nu$ with $\nu$ and $z$ being the correlation length exponent and dynamic exponent, respectively. The original KZ theory is established on the adiabatic-impulse-adiabatic picture, which typically requires an initial adiabatic stage~\cite{Kibble1976,Zurek1985}. Subsequent developments further generalized the KZ framework to protocols that start directly from the critical point, leading to the impulse–adiabatic scenario~\cite{polkovnikov2008prl,Polkovnikov2010prb,polkovnikov2010prb2,Dengshusa2011prb,Yin2016prb,nature2010fromcritical}.

On the other hand, for sudden quenches performed starting from the critical point $\lambda_c$ to a final value $\lambda_f$, Ref.~\cite{Polkovnikov2010prb} by De Grandi, Gritsev, and Polkovnikov demonstrates that the topological defect density $n$ follows
\begin{equation}
n\propto |g_f|^{d\nu},
\label{dgp}
\end{equation}
in which $g_f=\lambda_f-\lambda_c$. In the following, we refer to Eq.~(\ref{dgp}) as the De~Grandi-Gritsev-Polkovnikov (DGP) scaling. The DGP scaling arises from projecting the initial state onto the eigenstates of the Hamiltonian at $g_f$, thereby encoding the memory of the initial state. Moreover, such DGP scaling has been verified in exactly solvable models~\cite{Dengshusa2011prb}. Recently, it was shown that the DGP scaling, rather than the KZ scaling, can emerge for fast linear quench from the critical point when $R\gg|g_f|^{1/\nu r}$; In contrast, the KZ scaling relation of Eq.~(\ref{kz}) restores when $R\ll|g_f|^{1/\nu r}$~\cite{Zeng2023prl}. This phenomenon is then confirmed experimentally~\cite{Luole2025arxiv}.

However, these theories address distinct limiting cases. The KZ mechanism applies to slow near-adiabatic driving, whereas the DGP scaling applies to sudden quenches initiated at the critical point. To date, a unified scaling framework valid for arbitrary driving rates within the critical region remains absent.

To resolve these open issues, we systematically investigate the driven critical dynamics within the critical region. We find that scaling behaviors dominated by the critical point can emerge for arbitrary driving rates. Furthermore, we develop a generalized finite-time scaling (FTS) formalism that can fully capture the dynamic scenarios spanning various driving regimes. This framework not only unifies the KZ and DGP scaling limits, but also applies to initial states displaced from the critical point. We numerically verify the proposed scaling theory using the one-dimensional quantum Ising critical point and the Ising tricritical point, confirming the robustness and generality of the theory. Our results establish a unified description of distinct nonequilibrium dynamic regimes, deepening the fundamental understanding of how critical points govern driven dynamics under diverse driving protocols. It also paves the way for future investigations of the driven critical dynamics in more sophisticated theoretical models and experimental systems.

{\bf Upper bound on driving rate in original KZ theory}---In the original KZ framework, which focuses on the driven dynamics starting from an initial control parameter $\lambda_i$ sufficiently far from the critical point (typically assuming $\lambda_i \ll \lambda_c$). 
Under this condition, the driving rate $R$ cannot be arbitrarily large for the KZ scaling relation in Eq.~(\ref{kz}) to apply.

\begin{figure}[tbp]
\centering
  \includegraphics[width=0.9\linewidth,clip]{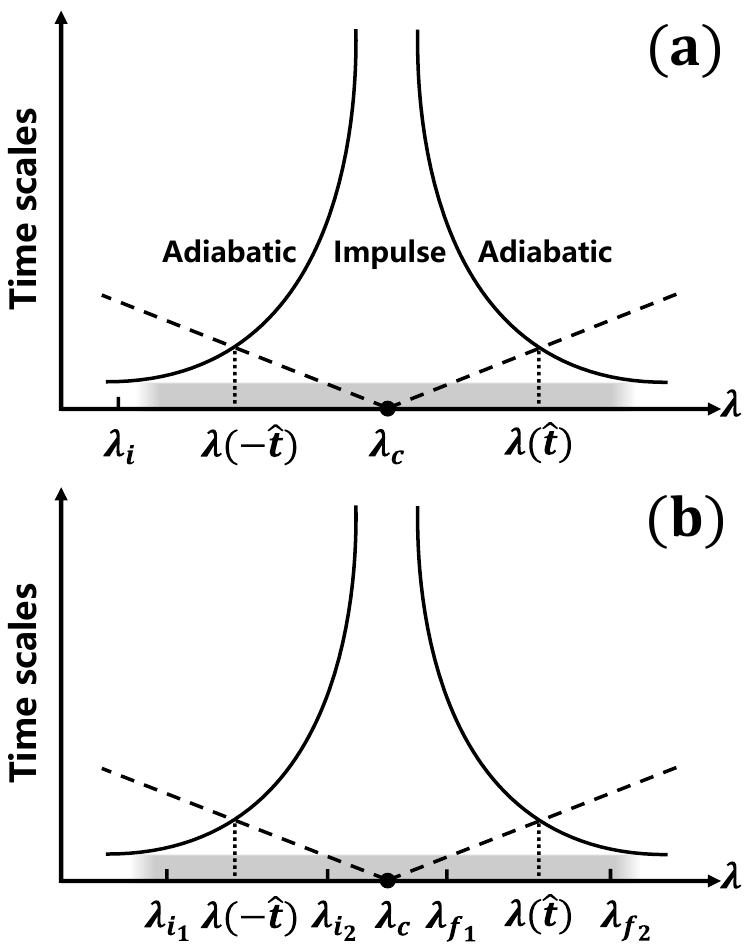}
  \vskip-3mm
  \caption{{\bf Time scales in driven critical dynamics.} The correlation time scale (solid curve) and the time distance to the critical point (dotted line) divide the dynamics into adiabatic and impulse regimes. (a) Conventional KZ dynamics starting far from the critical point requires the impulse regime to lie within the critical region, imposing an upper bound on the driving rate. (b) For driven dynamics initiated and evolving within the critical region, robust scaling holds for arbitrary driving rates and is described by the generalized FTS theory.
  }
  \label{fig:quench}
\end{figure}

To see this, we revisit the adiabatic-impulse-adiabatic scenario of the original KZ mechanism~\cite{Dziarmaga2010review,Polkovnikov2011rmp,delcamporev}, as illustrated in Fig.~\ref{fig:quench}(a). In the early stage of the ramp, the system evolves adiabatically, since the intrinsic relaxation time $\tau$ of the system is sufficiently short to allow the system to continuously track the instantaneous equilibrium state as $\lambda$ varies. As the system approaches the critical point, the equilibrium correlation time $\tau$ increases and eventually surpasses the time scale $|t|$ remaining until reaching $\lambda_c$. At the freeze-out time, defined by the condition $-\hat{t}\sim \hat{\tau}$, the system can no longer respond adiabatically to the ramp and enters the impulse regime, where its dynamics effectively ``freezes", until $\hat{t}\sim \hat{\tau}$ again on the other side of the critical point. The quantum KZ mechanism further posits that the topological defects induced by the driving are just contributed by the projection of the state at $-\hat{t}$ onto the excited states of the Hamiltonian at $\hat{t}$~\cite{Dziarmaga2010review,Polkovnikov2011rmp,Zoller2005prl,Dziarmaga2005prl}. 

For these topological defects to exhibit scaling behavior governed by the critical point, the impulse regime must lie within the critical region. This imposes the constraint $|\lambda(-\hat t)-\lambda_c|\leq|\lambda_G-\lambda_c|$, where $\lambda_G$ denotes the boundary of the critical region as determined by the Ginzburg criterion. Only in this way can the frozen state $|\psi(-\hat t)\rangle$ fully encode critical information. In addition, when $\lambda(-\hat t)$ is in the critical region, one has $|\lambda(-\hat t)-\lambda_c|\propto R^{1/\nu r}$. Accordingly, to ensure the validity of Eq.~(\ref{kz}), the driving rate $R$ must satisfy $R<R_G$, with the threshold rate $R_G$ obeying the scaling relation $R_G\sim|\lambda_G-\lambda_c|^{\nu r}$. Otherwise, non-universal fluctuation modes will enter the evolution of the impulse region, thereby breaking the scaling behavior~\cite{Silvi2016prl}.

As an extension of the KZ mechanism, the FTS theory has been established to describe scaling behaviors throughout the whole process in the critical region~\cite{Zhifangxu2005prb,Gong2010njp,Yin2014prb,huangyy2014prb,Feng2016prb}. Within this framework, the system is not completely frozen in the impulse regime; Rather, its dynamics is governed by the characteristic time scale $\hat{t}$, which dictates the dynamic scaling of all macroscopic quantities, ranging beyond the topological defects central to the original KZ picture~\cite{Zhifangxu2005prb,Gong2010njp}. Furthermore, the FTS can naturally accommodate other relevant quantities, such as symmetry-breaking field~\cite{huangyy2014prb,Yin2014prb,Feng2016prb,Yin2017prl}, temperature near quantum critical points~\cite{Yin2014prb}, and system size~\cite{huangyy2014prb,Yin2014prb}. It also applies to driving protocols beyond the conventional fluctuation-driven scenario~\cite{Gong2010njp,Yin2014prb}. Full scaling forms similar to the FTS have also been discussed from other settings~\cite{Deng2008epl,chandran2012prb,Polkovnikov2011prb,Huse2012prl,Liuchengwei2014prb}. These full scaling forms show great potential for applications in new quantum devices~\cite{Keesling2019,Ebadi2021,zhaihui2025,king2023nature,doi:10.1126/science.adx2728}. Here, since $\hat{t}$ can only reflect critical properties under the condition $R<R_G$, these full scaling theories also require an upper bound on the driving rate.

{\bf Generalized FTS form  with arbitrary driving rates}---In contrast, here we show that the physics becomes fundamentally different when both the initial and final quench parameters $\lambda_i$, $\lambda_f$, as well as the entire quench process, lie within the critical region. For such a setup, we find that universal scaling persists for arbitrary driving rates and can be consistently described by a generalized FTS form.

As illustrated in Fig.~\ref{fig:quench}(b), here the entire driven process is in the critical region. By incorporating all relevant parameters, we can obtain a generalized FTS form that describes the scaling behavior of a macroscopic quantity $P$ for arbitrary driving rates $R$ as
\begin{equation}
P[\lambda_i,\lambda(t),R]=R^{\kappa/r}f[g_iR^{-1/\nu r},g(t)R^{-1/\nu r},\{XR^{-x/r}\}],
\label{fts}
\end{equation}
in which $\kappa$ is the dimension of $P$, $g_i=\lambda_i-\lambda_c$, $g(t)=\lambda(t)-\lambda_c$, and $X$ and $x$ represent other possible relevant variables and their scaling dimensions. 

Different from the original FTS and other full scaling form~\cite{Zhifangxu2005prb,Gong2010njp,Yin2014prb,huangyy2014prb,Feng2016prb,Deng2008epl,chandran2012prb,Polkovnikov2011prb,Huse2012prl,Liuchengwei2014prb,Yin2017prl}, Eq.~(\ref{fts}) allows arbitrary driving rates with no upper bound and incorporates the initial parameters into the scaling function, offering a unified description on driven critical dynamics inside the critical region. In the following, we discuss different cases contained in Eq.~(\ref{fts}) in detail.

The simplest case in Eq.~(\ref{fts}) corresponds to $R<|g_i|^{\nu r}$ and $R<|g_f|^{\nu r}$. In this situation, the driven dynamics follows the standard adiabatic-impulse-adiabatic picture, as illustrated in Fig.~\ref{fig:quench}(b), where $\lambda_{i_1}$ and $\lambda_{f_2}$ denote the initial and final values of $\lambda$, respectively. In this way, $\hat t\propto R^{-z/r}$ is the only dominant time scale in the impulse region. Due to the initial adiabatic stage, the driven dynamics here can just be viewed as a segment of the original KZ dynamics starting far from the critical point. Accordingly, the initial parameter $\lambda_i$ (or equivalently $g_i$) can be regarded as residing at the infinite fixed points (corresponding to the completely ordered/disordered states) and thus exerts no explicit influence on the resulting scaling behavior. In this regime, Eq.~(\ref{fts}) reduces to the original FTS form with the term $g_i$ dropping out~\cite{Zhifangxu2005prb,Gong2010njp,Yin2014prb,huangyy2014prb,Feng2016prb}.

In contrast, when $R>|g_i|^{\nu r}$ (corresponding to $\lambda_{i_2}$ in Fig.~\ref{fig:quench}(b)), $\lambda_i$ already lies within the impulse regime. Under a crude approximation, the system prepared in the ground state at $\lambda_i$ exhibits frozen dynamics when driven near the critical point. Different values of $\lambda_i$ lead to distinct characteristic time scales. Accordingly, $\lambda_i$, or equivalently $g_i$, becomes a relevant parameter. Depending on the magnitude of $|g_f|^{\nu r}$ relative to $R$, two distinct dynamic scenarios emerge for the subsequent evolution governed by the final quench parameter $\lambda_f$:
\begin{enumerate}
    \item For $R<|g_f|^{\nu r}$ [corresponding to $\lambda_{f_2}$ in Fig.~\ref{fig:quench}(b)], dynamic freezing persists until $|g(t)|\sim R^{1/\nu r}$, marking the crossover threshold separating the impulse and adiabatic regimes.
    \item For $R>|g_f|^{\nu r}$ [corresponding to $\lambda_{f_1}$ in Fig.~\ref{fig:quench}(b)], frozen behavior prevails throughout the entire driving process, such that the evolution is primarily governed by the initial configuration.
\end{enumerate}
Both (1) and (2) differ markedly from the case with $R<|g_i|^{\nu r}$, where the influence of the initial parameter $\lambda_i$ (or equivalently $g_i$) becomes irrelevant. Here, by contrast, $\lambda_i$ (or $g_i$) and the associated initial state play an essential role in the subsequent evolution and must therefore be explicitly incorporated into the scaling form.

Physically, at slow driving rates, the system initialized in the ground state at $\lambda_i$ within the critical region generates only sparse low-energy excitations, whose behavior is governed by the critical point. At fast driving rates, the system retains memory of its initial state at $\lambda_i$. Since $\lambda_i$ lies already inside the critical region, the dynamics remain fully controlled by critical fluctuations. As a result, the universal properties of the driven dynamics inside the critical region are dictated by the critical point for arbitrary driving rates.

A special case for $R>|g_i|^{\nu r}$ is $\lambda_i=\lambda_c$ and $g_i=0$. Since $g_i$ is at its critical fixed point under this condition, it no longer appears in Eq.~(\ref{fts}). For the density of topological defects $n$, the parameter $\kappa$ in Eq.~(\ref{fts}) equals the spatial dimension $d$ of the system. If $R<|g(t)|^{\nu r}$, the scaling function $f$ converges to a constant, and the resulting $n$ obeys the KZ relation in Eq.~(\ref{kz})~\cite{Polkovnikov2010prb,Dengshusa2011prb,Zeng2023prl}. For $R>|g(t)|^{\nu r}$, $f$ scales as $[g(t)R^{-1/\nu r}]^{d\nu}$, leading to the DGP limit described by Eq.~(\ref{dgp})~\cite{Polkovnikov2010prb,Zeng2023prl}. 

As such, the generalized FTS form of Eq.~(\ref{fts}) not only unifies the two scaling limits but also accommodates initial states that deviate from the critical point, yielding a complete description of driven critical dynamics within the critical region.

\begin{figure}[tbp]
\centering
  \includegraphics[width=0.9\linewidth,clip]{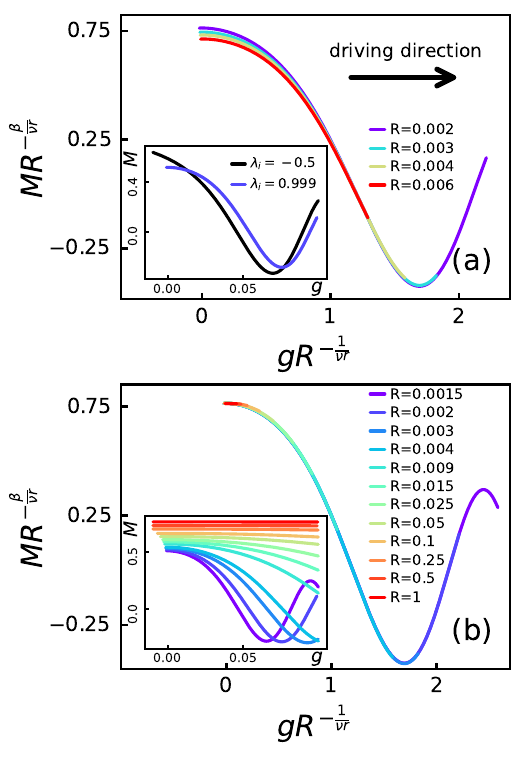}
  \vskip-3mm
  \caption{{\bf Driven critical dynamics for increasing $\lambda$ initiated from $\lambda_i$ close to the critical point.} (a) The original FTS fails to collapse data for the fixed initial parameters $\lambda_i=0.999$ and $h=0.00005$. The inset compares dynamic evolutions for different initial states at fixed $h=0.00005$ and $R=0.002$. (b) The generalized FTS gives excellent collapse over a wide range of $R$ for fixed $g_i R^{-1/\nu r}\approx-0.2236$ and $h R^{-\beta\delta/\nu r}\approx0.0167$.
  }
  \label{fig:ising1}
\end{figure}

{\bf Numerical results in critical point}--- To illustrate the scaling theory, we will
first take the $1$D quantum Ising model as an example. The Hamiltonian reads
\begin{equation}
H_I=-\sum_i \sigma^z_i \sigma^z_{i+1}-\lambda \sum_i \sigma^x_i-h \sum_i \sigma^z_i,
\label{ising}
\end{equation}
where $\sigma^{z,x}_i$ are the Pauli matrices in $z$- and $x$-directions, respectively, $\lambda$ is the transverse field and $h$ is the symmetry-breaking longitudinal field. The Ising coupling strength is set to be unity. The critical point of Eq.~(\ref{ising}) is $\lambda_{c}=1$. The exact critical exponents are $\beta=1/8$, $\nu=1$, $\delta=15$, and $z=1$. To ensure the entire process remains within the critical region, we set both $\lambda_i$ and $\lambda_f$ close to the critical point and take $|g_f|=0.1$.

The infinite time-evolving block decimation (iTEBD) algorithm~\cite{PhysRevLett.98.070201} is employed to compute the evolution of the system. The wave function is expressed as the matrix product state. The time evolution of the quantum state is implemented by updating the corresponding matrices via local evolution operators, which are constructed through the Suzuki–Trotter decomposition of $\exp(-iHt)$. The bond dimension in iTEBD is chosen as $100$, to which the numerical results are well converged.

We focus on the evolution of the order parameter $M= \langle\sigma_z\rangle$ for quenches initialized within the critical regime. To ensure a finite $M$, we introduce a weak external field $h$. According to Eq.~(\ref{fts}), $M$ should obey
\begin{equation}
M[\lambda_i,\lambda(t),R]=R^{\beta/\nu r}f_M[g_iR^{-1/\nu r},g(t)R^{-1/\nu r},hR^{-\beta\delta/\nu r}],
\label{ftsm}
\end{equation}
for arbitrary $R$.

As discussed above, the dynamics in the regime $R<|g_i|^{\nu r}$ recovers the original FTS and usual KZ scenario. We therefore focus on the case where $R>|g_i|^{\nu r}$. By increasing $\lambda$ to cross the critical point, the inset of Fig.~\ref{fig:ising1}(a) shows that the evolution of $M$ differs significantly between initial states near and far from the critical point, even $R$ and $h$ are the same for both cases, indicating that $R>|g_i|^{\nu r}$. Furthermore, rescaling data from the same initial state using the original FTS form, which does not include the initial information as the scaling variable, reveals that the rescaled curves fail to collapse, especially in the initial stage.

\begin{figure}[tbp]
\centering
  \includegraphics[width=0.9\linewidth,clip]{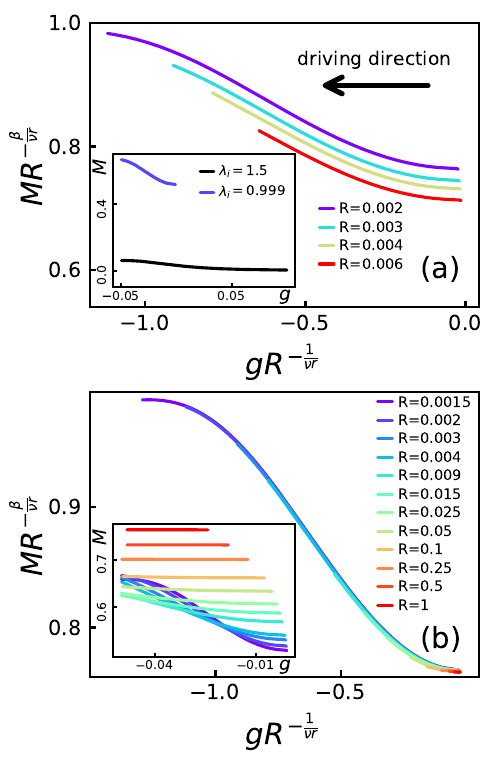}
  \vskip-3mm
  \caption{{\bf Driven critical dynamics for decreasing $\lambda$ initiated from $\lambda_i$ close to the critical point.} (a) The original FTS fails to collapse data for the fixed initial parameters for the fixed initial parameters $\lambda_i=0.999$ and $h=0.00005$. The inset compares dynamic evolutions for different initial states at fixed $h=0.00005$ and $R=0.002$. (b) The generalized FTS gives excellent collapse over a wide range of $R$ for fixed $g_i R^{-1/\nu r}\approx-0.2236$ and $h R^{-\beta\delta/\nu r}\approx0.0167$.
  }
  \label{fig:ising2}
\end{figure}

In contrast, Fig.~\ref{fig:ising1}(b) shows that for arbitrary fixed $g_i R^{-1/\nu r}\approx-0.2236$ and $h R^{-\beta\delta/\nu r}\approx0.0167$, the rescaled curves of $MR^{-\beta/\nu r}$ versus $g R^{-1/\nu r}$ collapse well onto each other over several orders of magnitude in $R$. These results verify Eq.~(\ref{ftsm}) and show that scaling remains present for $R>|g_i|^{\nu r}$, and the initial state must be rescaled to describe such scaling behaviors.

Moreover, for small $h$, the smallest value of $\lambda$ at which $M\approx 0$ can roughly represent the impulse-adiabatic boundary. Accordingly, as shown in Fig.~\ref{fig:ising1}(b), the driven dynamics for $R<0.009$ corresponds to the case of $R<|g_f|^{\nu r}$; while the driven dynamics for $R>0.009$ correspond to the case of $R>|g_f|^{\nu r}$. The larger $R$ is, the more pronounced the initial-state memory. When $R$ approaches unity, the value of $M$ remains nearly initial value throughout the entire evolution. The good scaling collapse in Fig.~\ref{fig:ising1}(b) demonstrates that all these behaviors can well be described by Eqs.~(\ref{fts}) and (\ref{ftsm}).

The generalized FTS form of Eq.~(\ref{fts}) can be further verified in the driving dynamics for decreasing $\lambda$. The inset of Fig.~\ref{fig:ising2}(a) illustrates that the dynamics initiated near the critical point differ distinctly from those starting far from the critical point. Simple rescaling according to the original FTS with fixed initial‑state parameters fails to collapse the corresponding curves in the whole stage. By contrast, as shown in Fig.~\ref{fig:ising2}(b), rescaled curves collapse onto a single universal profile for a broad range of $R$ when fixing $g_i R^{-1/\nu r}\approx-0.2236$ and $h R^{-\beta\delta/\nu r}\approx0.0167$, confirming the universality of Eq.~(\ref{fts}) and Eq.~(\ref{ftsm}).

{\bf Numerical results in tricritical point}--- In addition to usual critical point, we show that the generalized FTS is also applicable in driven dynamics at the tricritical point with arbitrary driving rates $R$. 

Recently, critical dynamics at tricritical points have attracted growing interest~\cite{YinJiang2025cpl,Lichengshu2025nc}, largely motivated by the search for exotic supersymmetric criticality in Rydberg atom systems~\cite{Guyingfei2014prl}. A typical model hosting the tricritical point is~\cite{Vishwanath2014science}
\begin{equation}
H=H_\sigma+H_\mu+H_{\sigma\mu},
\end{equation}
in which
\begin{subequations}
\begin{align}
    &H_\sigma=-\sum_i \sigma_i^z\sigma_{i+1}^z-\sum_i\sigma_i^x, \label{h2a} \\
    &H_\mu=\sum_i ( \mu_{i,a}^z\mu_{i,b}^z+ \mu_{i,b}^z\mu_{i+1,a}^z)-\lambda \sum_i (\mu_{i,a}^x+\mu_{i,b}^x) \\
    &-h\sum_i (\mu_{i,a}^z-\mu_{i,b}^z), \label{h2b} \\ &H_{\sigma\mu}=g\sum_i(\sigma_i^x\mu_{i,a}^z+\sigma_i^z\sigma_{i+1}^z\mu_{i,b}^z), \label{h2c}
\end{align}
\label{h2}%
\end{subequations}
where $\mu^{x,z}$ are Pauli matrices in $x$ and $z$ directions, $\lambda$ is the strength of the transverse fields, $h$ is the symmetry-breaking field, and $g$ is the coupling between two spin chains. When $g=0.5$, the tricritical point of this model is at $\lambda_c\approx1.27$~\cite{YinJiang2025cpl}. The order parameter is defined as $M=\langle(\mu_{i,a}^z-\mu_{i,b}^z)/2\rangle$. The critical exponents along the $\lambda$-direction are $\beta=1/4$, $\nu=5/4$, $\delta=9$, and $z=1$~\cite{Vishwanath2014science}.

\begin{figure}[tbp]
\centering
  \includegraphics[width=0.9\linewidth,clip]{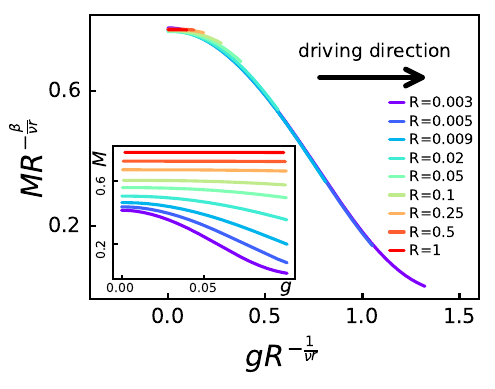}
  \vskip-3mm
  \caption{{\bf Driven critical dynamics for increasing $\lambda$ initiated from the tricritical point.} For fixed $g_i=0$ and $h R^{-\beta\delta/\nu r}\approx0.1667$, rescaled curves of $M$ versus $g$ collapse well for a wide range of $R$ according to the generalized FTS.
  }
  \label{fig:susy}
\end{figure}

We investigate driven dynamics starting from $\lambda_i=\lambda_c$. As illustrated in Fig.~\ref{fig:susy}, the rescaled curves of $MR^{-\beta/\nu r}$ against $g R^{-1/\nu r}$ exhibit good collapse over a broad range of $R$, with $hR^{-\beta\delta/\nu r}$ fixed at approximately $0.1667$ (here $g_i=0$). Our findings demonstrate that driven dynamic scaling behavior persists for arbitrary driving rates in the critical region of the tricritical point and the critical dynamics are well captured by Eqs.~(\ref{fts}) and (\ref{ftsm}), solidifying the universality of our theoretical framework.

{\bf Summary and Discussion}---In this paper, we develop a generalized FTS theory that provides a unified description of driven critical dynamics within the critical region for arbitrary driving rates. This framework not only unifies the slow-driving KZ regime and the sudden quench DGP limit, but also accommodates initial states that deviate from the critical point. Numerical simulations at both quantum critical and tricritical points validate the generalized FTS scaling form, confirming its robustness and broad applicability. Owing to its inherent universality, we expect this theory to be extendable to other critical points, including those beyond the Landau paradigm~\cite{Shu20231,zengFTSKZM2025,Yin2020PhysRevResearch,PhysRevA.101.023610,Yinarxiv2024,scjr-r3hj,qtwr-8kth}. It also applies to critical dynamics under alternative driving protocols, such as those with varying symmetry-breaking fields~\cite{Gong2010njp,Yin2017prl} and changing temperature near quantum critical point~\cite{Yin2014prb}.

Our theory is immediately relevant to state-of-the-art experiments in fast-developing quantum devices, where the original KZ mechanism and the original FTS has been realized with high controllability~\cite{Keesling2019,Ebadi2021,zhaihui2025,king2023nature,doi:10.1126/science.adx2728}. By removing the rate constraint of the conventional KZ framework, our results allow reliable characterization of universal nonequilibrium critical behaviors when quenches start inside the critical region. Our unified framework offers a general foundation for characterizing universal nonequilibrium critical behaviors in broad quantum quench experiments.

{\bf Acknowledgments}---We thank Zhi-Xuan Li and Hua-Bi Zeng for helpful discussions. This project is supported by the National Natural Science Foundation of China (Grants No. 12222515), the Research Center for Magnetoelectric Physics of Guangdong Province (Grant No. 2024B0303390001), the Guangdong Provincial Key Laboratory of Magnetoelectric Physics and Devices (Grant No. 2022B1212010008), the Science and Technology Projects in Guangzhou City (Grant No. 2025A04J5408), and Quantum Science and
Technology-National Science and Technology Major Project
(Grant No.2025ZD0300400).


\bibliographystyle{apsrev4-2}
\let\oldaddcontentsline\addcontentsline
\renewcommand{\addcontentsline}[3]{}
\bibliography{ref.bib}
\let\addcontentsline\oldaddcontentsline

\end{document}